\documentclass[twocolumn,preprintnumbers,superscriptaddress,nofootinbib,aps,prl,floatfix]{revtex4}

\usepackage{graphicx}  
\usepackage{dcolumn}   
\usepackage{bm}        
\usepackage{amssymb,amsmath}
\usepackage{wasysym}
\usepackage{adjustbox,lipsum}
\usepackage{subfigure}

\def\beq{\begin{equation}}
\def\eeq{\end{equation}}
\usepackage[usenames,dvipsnames,svgnames]{xcolor}
\usepackage{hyperref}
\hypersetup{colorlinks, citecolor=black, linkcolor=black, urlcolor=black}

\usepackage{textcomp}
\usepackage{amsmath}
\usepackage{amsfonts}
\usepackage{amssymb}
\usepackage{bbm}                
\usepackage{graphicx, rotating}
\usepackage{epstopdf}
\usepackage{epsfig}
\usepackage{latexsym}
\usepackage{graphicx}
\usepackage{bbold}
\usepackage{color}
\usepackage{amsmath,amssymb}
\usepackage{slashed}

\newcommand{\bea}{\begin{eqnarray}}
\newcommand{\eea}{\end{eqnarray}}
\newcommand{\eq}[1]{Eq.~(\ref{#1})}

\def\stw{s_{\theta_W}}
\def\ctw{c_{\theta_W}}
\def\ttw{t_{\theta_W}}
\def\lra#1{\overset{\text{\scriptsize$\leftrightarrow$}}{#1}}

\preprint{IPPP/18/53}

\begin{document}
\title{Probing Electroweak Precision Physics via \\ boosted Higgs-strahlung  at the LHC}
\author{Shankha Banerjee}\email{shankha.banerjee@durham.ac.uk}
\affiliation{Institute of Particle Physics Phenomenology, Durham University, Durham DH1 3LE, UK}
\author{Christoph Englert}\email{christoph.englert@glasgow.ac.uk}
\affiliation{SUPA, School of Physics and Astronomy, University of Glasgow, Glasgow G12 8QQ, UK}
\author{Rick S. Gupta}\email{sandeepan.gupta@durham.ac.uk}
\affiliation{Institute of Particle Physics Phenomenology, Durham University, Durham DH1 3LE, UK}
\author{Michael Spannowsky}\email{michael.spannowsky@durham.ac.uk}
\affiliation{Institute of Particle Physics Phenomenology, Durham University, Durham DH1 3LE, UK}

\begin{abstract}
We study the process $pp \to Z(\ell^+ \ell^-)h(b\bar b)$ in the Standard Model Effective Field Theory (SMEFT) 
at high energies using subjet techniques to reconstruct the Higgs boson. We show that  at high 
energies this process probes four directions in the dimension 6 EFT space, namely the operators that 
contribute to the four contact interactions, $hZ_\mu \bar{f}\gamma^\mu f$, where $f=u_L, u_R,d_L$ and 
$d_R$. These four directions are, however, already constrained by the $Z$-pole and diboson 
measurements at LEP.  We show that by utilising the energy growth of this process in the SMEFT and 
the accuracy that can be achieved by using subjet techniques at the High Luminosity LHC, one can 
obtain bounds on these operators that are an order of magnitude better than existing LEP bounds.
\end{abstract}

\maketitle

\subsection{Introduction}
Characterizing the properties of the Higgs boson is arguably the most concrete particle physics goal 
of our time. This is further motivated by the dearth of any signs of physics beyond the Standard Model 
(BSM) in LHC data so far. One well-motivated course of action in this situation is to probe heavy 
new physics outside the reach of direct searches via precise indirect measurements. A historic example 
of constraining high energy physics even beyond the energy coverage of a collider is the LEP experiment, 
which was able to probe scales up to the few TeV via indirect precision measurements although it ran at 
a much smaller collision energy.  

As the Higgs boson could not be produced before the LHC experiment under controlled conditions, one 
might naively think that any measurement of interactions involving the Higgs boson is complementary 
to past measurements. However, an Effective Field Theory (EFT) perspective allows us to correlate 
measurements at different energy scales only on the basis of SM symmetry and matter content. In fact, 
there are fewer $SU(2)_L\times U(1)_Y$ invariant operators at the lowest order in the LHC-relevant 
EFT expansion at dimension 6~\cite{Grzadkowski:2010es} than the number of (pseudo-)observables they 
contribute to. As a result, correlations between LHC and LEP measurements can be exploited. For 
instance the LEP measurements of $Z$-boson pole observables and anomalous Triple Gauge Couplings (TGCs)  inform the measurement of Higgs observables at the LHC as they can be related 
to a common gauge-invariant set of SMEFT operators. The analysis in Ref.~\cite{Gupta:2014rxa} reveals, 
in fact, that apart from eight Higgs observables, the so called ``Higgs primaries'', all other Higgs 
interactions present in the dimension-6 Lagrangian can be constrained already by $Z$-pole and diboson 
measurements at LEP. 

For the set of already constrained Higgs coupling deformations, the LHC has to compete with LEP's 
precision to add new information in order to gain a more complete picture. This might seem challenging 
given that the LHC is intrinsically less accurate compared to LEP. The key advantage of the LHC (and other 
future colliders), however, is a much larger energy reach compared to LEP, thus allowing us to constrain 
new physics from a plethora of available kinematical information~(see 
also~\cite{Banerjee:2013apa,Amar:2014fpa,Buschmann:2014sia,Craig:2014una, Banerjee:2015bla,Cohen:2016bsd,Ge:2016zro,Denizli:2017pyu,Barklow:2017suo,Barklow:2017awn,Khanpour:2017cfq}). As we will see, the high energy sensitivity of the LHC will allow us to strongly constrain EFT-induced anomalous couplings involved in processes that grow with energy:
\beq
\label{eq:growth}
  \frac{\delta\sigma (\hat{s})}{\sigma_{SM} (\hat{s})}\sim \delta g_i    \frac{\hat{s}}{m_Z^2}\,.
\eeq
From Eq.~\eqref{eq:growth}, we see that the anomalous coupling $g_i$ can be measured/constrained at 
the per-mille to percent level even if the underlying sensitivity to the fractional cross-section 
deviation, $\delta\sigma/\sigma_{SM}$ is only  ${\cal O}(30\%)$ at high energies.  

The specific process we are interested in here is Higgs-strahlung, $pp \to Z(\ell^+ \ell^-)h(b\bar 
b)$. Studying the $h \to b\bar b$ mode instead of the $h \to \gamma \gamma$ leads to a big enhancement 
in the rate but the Higgs-strahlung process still remains challenging with an ${\cal O}(50)$ background-to-signal
ratio. Relating such a systematics limited result to the extraction of Higgs couplings can be at odds 
with the implicit assumption of perturbativity of the  EFT expansion. We technically rely on the latter
to perform proof-of-principle analyses and eventually full searches at ATLAS and CMS. As $\delta\sigma/\sigma_{SM} \gtrsim 1$ signals the 
breakdown of EFT validity for weakly coupled UV completions~\cite{Contino:2016jqw},  a sensitivity 
to  smaller values of $\delta\sigma/\sigma_{SM}$ is essential. To gain such 
precision, we need high luminosities (at least 300 fb$^{-1}$) and  advanced boosted Higgs tagging 
techniques which can reduce the ratio of the number of $Zb\bar b$ to the SM $Zh(b\bar b)$ events to 
an ${\cal O}(1)$ number as shown earlier in Refs.~\cite{Butterworth:2008iy, Soper:2010xk, Soper:2011cr}. 
This work is, therefore, an example of a study at the ``high energy-luminosity'' frontier in the 
spirit of Ref.~\cite{deBlas:2013qqa,Farina:2016rws,Franceschini:2017xkh}. 

While adding the channel $pp \to Z(\nu \bar{\nu})h(b\bar 
b)$  can further improve the limits on the effective operators we study \cite{Butterworth:2008iy}, this channel is subjected to backgrounds and employs observables with larger systematic uncertainties. We therefore leave an inclusion of this channel for future work.


As we will see, the leading high energy contribution to the $pp \to Zh$ process comes from the four 
contact interactions  $hZ_\mu \bar{u}_{L,R} \gamma^\mu u_{L,R}$ and $hZ_\mu \bar{d}_{L,R} \gamma^\mu 
d_{L,R}$ that are present in the dimension-6 extended Lagrangian. Thus, although many more operators 
contribute to the  $pp \to Zh$ process, the high energy limit isolates the four linear combinations of 
operators that generate the above contact terms. An interesting observation, first made in 
Ref.~\cite{Franceschini:2017xkh}, is that the same four EFT directions (that the authors call ``high 
energy primaries'') also control $Wh$ and $WW/WZ$ production. The reason is that at high energies these 
four final processes  correspond to the production of different components of the Higgs doublet due 
to the Goldstone Boson Equivalence Theorem~\cite{Chanowitz:1985hj}. They are therefore related by 
$SU(2)_L$ symmetry for $\hat{s}\gg m^2_Z$. Hence, although these four diboson processes may be very 
different from a collider physics point of view, they are intimately related by gauge symmetry, which 
stands at the heart of an EFT interpretation.  This enables an elegant understanding of the connection of pseudo-observables in $WW$ production 
(such as TGCs) with those in $Zh$ production. It will also allow us to present our results in a 
combined way with the projections for $WZ$ production in Ref.~\cite{Franceschini:2017xkh}.

\begin{table}[t]
\begin{center}
\begin{adjustbox}{width=\columnwidth}
\small
\begin{tabular}{c|c}
SILH Basis&Warsaw Basis\\\hline
\rule[-1.2em]{0pt}{3em}$\displaystyle{\cal O}_W=\frac{ig}{2}\left( H^\dagger  \sigma^a \lra {D^\mu} H \right )D^\nu  W_{\mu \nu}^a$&$\displaystyle{\cal O}^{(3)}_L= (\bar{Q}_L \sigma^a\gamma^\mu Q_L)(i H^\dagger \sigma^a\lra D_\mu   H)$\\
\rule[-1.2em]{0pt}{3em}$\displaystyle{\cal O}_B=\frac{ig'}{2}\left( H^\dagger  \lra {D^\mu} H \right )\partial^\nu  B_{\mu \nu}$&$\displaystyle{\cal O}_L=  (\bar{Q}_L \gamma^\mu Q_L)(i H^\dagger \lra D_\mu   H)$\\
\rule[-.6em]{0pt}{1.5em}$\displaystyle{\cal O}_{HW}=i g(D^\mu H)^\dagger\sigma^a(D^\nu H)W^a_{\mu\nu}$&
$\displaystyle{\cal O}^u_R= (\bar{u}_R \gamma^\mu u_R)(i H^\dagger \lra D_\mu   H)$\\
\rule[-.6em]{0pt}{1.5em}$\displaystyle{\cal O}_{HB}=i g'(D^\mu H)^\dagger(D^\nu H)B_{\mu\nu}$&$\displaystyle{\cal O}^d_R=  (\bar{d}_R \gamma^\mu d_R)(i H^\dagger \lra D_\mu   H)$\\
\rule[-1.2em]{0pt}{3em}$\displaystyle{\cal O}_{2W}=-\frac{1}{2}  ( D^\mu  W_{\mu \nu}^a)^2$&\\
\rule{0pt}{1.5em}$\displaystyle{\cal O}_{2B}=-\frac{1}{2}( \partial^\mu  B_{\mu \nu})^2$&
 \end{tabular}
 \end{adjustbox} 
  \caption{Dimension-six operators that give dominant contribution to $pp \to Vh$ at high energies 
  in the Warsaw~\cite{Grzadkowski:2010es} and SILH~\cite{Giudice:2007fh} bases. \label{list}}
\end{center}
\end{table}

\begin{table*}[!t]
\begin{center}
\small
\begin{tabular}{|c|c|}
\hline
&EFT directions probed by high energy $ff \to Vh$ production \\
\hline
\hline
 Warsaw Basis~\cite{Grzadkowski:2010es} & $- \frac{2 g}{\ctw}\frac{v^2}{\Lambda^2}(|T_3^f|c^{1}_L-T_3^f c^{3}_L+(1/2-|T_3^f|)c_f)$\\
BSM Primaries~\cite{Gupta:2014rxa} &$   \frac{2 g}{\ctw}Y_f \ttw^2 \delta \kappa_\gamma+2 \delta g^Z_{f}- \frac{2 g}{\ctw}(T^f_3 \ctw^2 + Y_f \stw^2)\delta g_1^Z $\\
 SILH Lagrangian~\cite{Giudice:2007fh} &$  \frac{g}{\ctw}\frac{m_W^2}{\Lambda^2}(2 T_3^f\hat{c}_W-2\ttw^2 Y_f \hat{c}_B)$\\
 Universal observables &$ \frac{2 g}{\ctw}Y_f \ttw^2  (\delta \kappa_\gamma-\hat{S}+Y)- \frac{2 g}{\ctw}(T^f_3 \ctw^2 + Y_f \stw^2)\delta g_1^Z- \frac{2 g}{\ctw}T^f_3 W$\\
 High Energy Primaries~\cite{Franceschini:2017xkh} &$ -\frac{2 m_W^2}{g \ctw }(|T_3^f| a_q^{(1)}-T_3^f a_q^{(3)}+(1/2-|T_3^f|)a_f)$\\
\hline
 \end{tabular}
  \caption{ The linear combinations of Wilson coefficients contributing to the contact interaction couplings $g^h_{Zf}$ that control the $ff \to Vh$ process at high energies. The four directions relevant for hadron colliders (corresponding to $f=u_L, d_L, u_R, d_R$)  can  be read off from this table by substituting the value of the $SU(2)_L$ and $U(1)_Y$ quantum numbers  $T_3^f$ and $Y_f$ for the corresponding $f$. Here $\hat{c}_W=c_W+c_{HW}-c_{2W}$ and $\hat{c}_B= c_B+c_{HB}-c_{2B}$. For the nomenclature of the operators, their corresponding Wilson coefficients and observables see for eg. Ref.~\cite{Franceschini:2017xkh}.}
  \label{dirn}
\end{center}
\end{table*}

\subsection{The high energy $Vh$-amplitude in the SMEFT}
Let us first study $Vh$ production at high energy in the SMEFT where $V=W,Z$. Although we focus on 
$pp \to Zh$ production in the subsequent sections, here we keep the discussion more general 
considering also the $Wh$ final state. We will see that $Vh$ production at 
hadron colliders at high energies, isolates four independent directions in the full 59 dimensional 
space of dimension 6 operators.  To derive this fact, consider first the vertices in the dimension 6 Lagrangian that contribute 
to the $ff \to Zh$  process in unitary gauge, 
\begin{multline}
\Delta {\cal L}_6\supset \sum_f {\delta g^Z_{f}} Z_\mu \bar{f} \gamma^\mu f +\delta g^W_{ud} (W^+_\mu \bar{u}_L \gamma^\mu d_L+h.c.)\\
+ {g^h_{VV}}\,  h \left[W^{+\, \mu} W^-_\mu+\frac{1}{2\ctw^2}Z^\mu Z_\mu\right]
+  \delta g^h_{ZZ}\, h \frac{Z^\mu Z_\mu}{2\ctw^2}\\ 
+ \sum_f g^h_{Zf}\,\frac{h}{v}Z_\mu \bar{f} \gamma^\mu f+g^h_{Wud}\,\frac{h}{v}(W^+_\mu \bar{u}_L \gamma^\mu d_L+h.c.)\\
+{\kappa_{Z \gamma }}\frac{h}{v} A^{\mu\nu} Z_{\mu\nu}+\kappa_{WW}\,\frac{h}{v}
 W^{+\, \mu\nu}W^-_{\mu\nu}
+\kappa_{ZZ}\,\frac{h}{2v} Z^{\mu\nu}Z_{\mu\nu}\,.
\label{lagr}
\end{multline}
We are using the formalism presented in Ref.~\cite{Gupta:2014rxa, Pomarol:2014dya} where $\alpha_{em}$, $m_Z$ and $m_W$   have been used as input parameters and any corrections 
to the SM vector propagators, \textit{i.e.} the terms $V_\mu V^\mu, V_{\mu\nu} V^{\mu\nu}$ and  $V_{\mu\nu} F^{\mu\nu}$, have 
been traded in favor of the vertex corrections. Note that the above parameterisation is equivalent to the one in Ref.~\cite{Gonzalez-Alonso:2014eva, Greljo:2015sla} (see Ref.~\cite{Greljo:2017spw}).   Keeping only the leading terms in $\hat{s}/m_Z^2$ in the BSM correction, we obtain for the amplitude $ \mathcal{M}(ff\to V_{T,L} h)$,
\begin{equation}
  \label{amps}
\begin{split}
 Z_T h&: \, {g^Z_{{f}}} \frac{ \epsilon^*\cdot J_{f}}{v}  \frac{2  m_Z^2}{\hat{s}}\  \,  \, \Bigg[1+ \left(\frac{g^h_{Zff}}{g^Z_f}-  \kappa_{ZZ}\right)  \frac{\hat{s}}{2 m_Z^2}  \Bigg] \, ,\\
Z_L h&: \, {g^Z_{{f}}} \frac{ q\cdot J_{f}}{v}  \frac{2  m_Z}{\hat{s}}\  \,  \, \Bigg[1+ \frac{{g}^h_{Zff}}{g^Z_{f}} \frac{\hat{s}}{2 m_Z^2} \Bigg] \,,\\
  W_T h  &: \, {g^W_{{f}}} \frac{ \epsilon^*\cdot J_{f}}{v}  \frac{2  m_W^2}{\hat{s}}\  \,  \, \Bigg[1+\left(\frac{g^h_{Wff'}}{g^W_f}-  \kappa_{WW}\right)  \frac{\hat{s}}{2 m_W^2}  \Bigg] \, ,\\
  W_L h  &: \,  g^W_{f}  \frac{ q\cdot J_{f}}{v}  \frac{2  m_W}{\hat{s}}\  \,  \, \Bigg[1+\frac{g^h_{Wff'}}{g^W_f} \frac{\hat{s}}{2 m_W^2} \Bigg] \,, 
\end{split}
\end{equation}
 where $g^Z_{f}=g(T_3^f-Q_f \stw^2)/\ctw$, and $g^W_f={g}/{\sqrt{2}}$. $J^\mu_f$ is the fermion current $\bar{f} \gamma^\mu f$, the subscript $L$ ($T$) denotes the longitudinal (transverse) polarization of the gauge boson, $q$ denotes its four momentum and $\epsilon$ the   polarization vector. 
 
We see that only the ${g}^h_{Vf}$ and $\kappa_{VV}$ couplings lead to an amplitude 
growing with energy. In the case of the $\kappa_{VV}$ couplings, the energy growth arises because of 
the extra powers of momenta in the $hVV$ vertex, whereas for the contact interaction, ${g}^h_{Vf}$, 
the energy growth is due to the fact that there is no propagator in the diagram involving this vertex.
In fact for the latter interaction, the only difference in the amplitude with respect to the SM is 
the absence of the propagator. Thus, angular distributions 
are expected to be identical for BSM and SM production. Therefore, the only way to probe this interaction is through the 
direct energy-dependence of differential cross-sections.

On the other hand, the $\kappa_{VV}$ interactions contribute only to the transverse $V$ amplitude as 
a consequence of their vertex structure. Hence, they cannot interfere with the dominant longitudinal 
piece in the SM amplitude. As a result, after summing over all $V$-polarizations, the leading 
piece in the high energy cross-section deviation, is controlled only by the couplings ${g}^h_{Vf}$ 
whereas the $\kappa_{VV}$ contribution is suppressed by an additional ${\cal O} (m_V^2/\hat{s})$ 
factor. 

Note that the couplings, $\delta g^Z_{f}$ and $\delta g^h_{ZZ}$ also lead to deviations from the SM amplitude but these corrections do not grow with energy and are also suppressed by an additional ${\cal O} (m_V^2/\hat{s})$ 
factor with respect to the ${g}^h_{Vf}$ contribution. We have checked explicitly that including these couplings have no noticeable impact on our analysis.

At hadron colliders, the $pp \to Vh$ process at high energies and at leading order are therefore 
controlled by the five contact interactions:  $g^h_{Zf}$, with $f=u_L, u_R,d_L$ and $d_R$ and 
$g^h_{Wud}$. These five couplings correspond to different linear combinations of Wilson coefficients 
in any given basis. In Tab.~\ref{list} we show all operators in the ``Warsaw''~\cite{Grzadkowski:2010es} 
and strongly-interacting light Higgs (SILH)~\cite{Giudice:2007fh} bases that generate these contact 
terms. As there are only four independent operators contributing to these five interactions in the 
Warsaw basis, there exists a basis independent constraint at the dimension-6 level,
\beq
g^h_{Wud}=\ctw\frac{g^h_{Zu_L}-g^h_{Zd_L}}{\sqrt{2}}
\eeq
leaving only the four independent $g^h_{Zf}$ couplings. 

In Table~\ref{dirn}, we show the linear combinations of Wilson coefficients contributing to the four 
$g^h_{Zf}$ couplings in different EFT bases. The first row gives these directions in the Warsaw basis. 
The second row provides the aforementioned directions in the BSM Primary basis of Ref.~\cite{Gupta:2014rxa}, 
where the Wilson coefficients can be written in terms of already constrained pseudo-observables.  It is clear in this basis that the directions to be probed by high energy $Vh$ production can be 
written in terms of the LEP (pseudo)observables. The couplings $\delta g^Z_{f}$ defined in \eq{lagr} 
are strongly constrained by  $Z$-pole measurements at LEP,  whereas the anomalous TGCs, $\delta \kappa_\gamma$ and 
$\delta g_1^Z$ (in the notation of Ref.~\cite{Hagiwara:1986vm}),   were constrained by $WW$ 
production during LEP2. 

For the physically motivated case where the leading effects of new physics 
can be parametrized by universal (bosonic) operators, the SILH Lagrangian provides a convenient 
formulation and we show the above directions in this basis in the third row of Table~\ref{dirn}. For this case, as
shown in the fourth row of Table~\ref{dirn}, one 
can again write the directions in terms of only the ``oblique''/universal pseudo-observables, 
\textit{viz.}, the TGCs $\delta \kappa_\gamma$ and $\delta g_1^Z$ and the Peskin-Takeuchi 
$\hat{S}$-parameter~\cite{Peskin:1991sw} in the normalization of Ref.~\cite{Barbieri:2004qk}. For a definition of these observables we refer to the Lagrangian presented in Ref.~\cite{Elias-Miro:2013eta} (see also Ref.~\cite{Wells:2015uba}).  As we already mentioned, upon using the Goldstone 
Equivalence Principle, one finds that the same 4 dimensional subspace of operators also controls the 
longitudinal $VV$ production at high energies. This space is defined in Ref.~\cite{Franceschini:2017xkh} 
in terms of the four high energy primaries which are linear combinations of the four ${g}^h_{Vf}$ 
couplings, as shown in the last row of Table~\ref{dirn}.

As it is not possible to control the polarization of the initial state partons in a hadron collider, 
the process can, in reality, only probe two of the above four directions. Taking only the interference 
term, these directions are 
\begin{equation}
\begin{split}
g^Z_{\textbf{u}}&= g^h_{Zu_L}+\frac{g^Z_{u_R}}{g^Z_{u_L}} g^h_{Zu_R}\,, \\
g^Z_{\textbf{d}}&= g^h_{Zd_L}+\frac{g^Z_{d_R}}{g^Z_{d_L}} g^h_{Zd_R}\,,
\end{split}
\end{equation}
 where $g^Z_f$ is defined below \eq{amps}. Also, at a given energy, the interference term for the 
 $pp \to Zh$ process is sensitive only to a linear combination of the up-type and down-type coupling 
 deviations, \textit{i.e.}, to the direction, 
\beq
g^Z_{\textbf{p}}= g^Z_{\textbf{u}}+ \frac{{\cal L}_d(\hat{s})}{{\cal L}_u(\hat{s})}g^Z_{\textbf{d}}
\eeq 
where ${\cal L}_{u,d}$ is the $u \bar{u}$, $d \bar{d}$ luminosity at a given partonic centre of mass 
energy. We find that the luminosity ratio changes very little with energy (between 0.65 and 0.59 if 
$\sqrt{\hat{s}}$ is varied between 1 and 2 TeV). Thus, to a good approximation, $pp \to Zh$ probes 
the single direction in EFT space given by 
\beq
g^Z_{\textbf{p}}=g^h_{Zu_L} -0.76~g^h_{Zd_L}   - 0.45~g^h_{Zu_R} + 0.14~g^h_{Zd_R}  \,,
\label{compdir}
\eeq 
where we have substituted the values for $g^Z_f$ and evaluated the luminosities at the energy 
${\hat{s}}=(1.5~\text{TeV})^2$. This can now be written in terms of the LEP-constrained pseudo-observables in the second and fourth row of Tab.~\ref{dirn},
\bea
g^h_{Z\textbf{p}}&=&2~\delta g^h_{Zu_L} -1.52~\delta g^h_{Zd_L}   - 0.90~\delta g^h_{Zu_R} + 0.28~\delta g^h_{Zd_R}\nonumber\\
&&-0.14~\delta \kappa_\gamma   -0.89~\delta g^Z_1 \nonumber\\
g^h_{Z\textbf{p}}&=&-0.14~(\delta \kappa_\gamma-\hat{S}+Y) -0.89~\delta g^Z_1  -1.3~W
\label{diru}
\eea
where the first line applies to the general case and the second line to the universal case.

Note that in the discussion so far we have not considered the $gg \to Zh$ production channel at 
hadron colliders~\cite{Dicus:1988yh,Kniehl:1990iva,Matsuura:1990ba,Kniehl:2011aa,Harlander:2013mla,Englert:2013vua,Altenkamp:2012sx,Hespel:2015zea,Harlander:2018yns}. 
While formally a higher order correction, after all the cuts are applied, this subprocess contributes 
an appreciable 15\% of the total SM $pp \to Zh$ cross-section in our analysis due to the top-threshold 
inducing boosted final states~\cite{Englert:2013vua}. We find, however, that introduction of the EFT 
operators does not lead to a energy growing amplitude with this initial state, and thus this channel 
has a subdominant contribution to the EFT signal. While we fully include this contribution in our 
collider analysis, the introduction of this channel does not alter the discussion so far in an 
important way.

 
\medskip
 
We now turn to the crucial issue of estimating the scale of new physics (and thus the cut-off for 
our EFT treatment) for a given size of the couplings, ${g}^h_{Vf}$. This will also give us an idea of 
the new physics scenarios that our analysis can probe. As is clear from the operators in Tab.~\ref{list}, 
the ${g}^h_{Vf}$ couplings arise from current-current operators that can be generated, for instance, 
by integrating out at tree-level a heavy $SU(2)_L$ triplet (singlet) vector $W'^a$ ($Z'$) that couples 
to SM fermion currents, $\bar{f} \sigma^a \gamma_\mu f$ ($\bar{f} \gamma_\mu f)$ with a coupling 
$g_f$ and to the Higgs current $i{H}^\dagger \sigma^a \lra{D}_\mu H$ ($i{H}^\dagger \lra{D}_\mu H$) 
with a coupling $g_H$, 
\beq
{g}^h_{Zf}\sim \frac{g_H g g_f v^2}{\Lambda^2}\,,
\eeq
where $\Lambda$ is the mass of the vector and therefore the matching scale or cut-off of the low 
energy EFT. The coupling to the SM fermions can be universal if the heavy vector couples to them only 
via kinetic mixing with the SM gauge bosons. This results in a coupling of the heavy vector to the 
$SU(2)_L$ and hypercharge currents given by $g_W=g/2$ and $g_B=g' Y_f$, $Y_f$ being the SM hypercharge. 
As we want our results to be applicable to the universal case, we assume the coupling $g_f$ to be 
given by a combination of $g_B$ and $g_W$ to obtain,
\begin{equation}
  \label{cutoff}
 \begin{split}
{g}^h_{Zu_L,d_L} &\sim \frac{g_H g^2  v^2}{ 2\Lambda^2}\,,\\
{g}^h_{Zu_R,d_R} &\sim \frac{g_H g g' Y_{u_R,d_R} v^2}{\Lambda^2} \,,
\end{split}
\end{equation}
 and then further assume a weakly coupled scenario with $g_H=1$ (note that this is a bit larger than the corresponding value $g_H=g/(2 \ctw)$ for the SM  $hZZ$ coupling). In the above equation, we have 
 ignored the smaller contributions from $g_B$ to the left-handed couplings. For any set of couplings 
 $\{{g}^h_{Zu_L} ,{g}^h_{Zd_L} ,{g}^h_{Zu_R} ,{g}^h_{Zd_R} \}$, we evaluate the cut-off using 
 \eq{cutoff} with $g_H=1$ and take the smallest of the four values. It is clear that for strongly 
 coupled scenarios with larger values of $g_H$, the cut-off assumed in our analysis is smaller than 
 necessary and thus our projected bounds will be conservative.

\begin{table*}[ht]
\begin{center}
\small
\begin{tabular}{||c|c|c||}
\hline
Cuts &  $Zb\bar{b}$ & $Zh$ (SM) \\
\hline \hline
At least 1 fat jet with 2 $B$-mesons with $p_T > 15$ GeV & 0.23 & 0.41 \\
2 OSSF isolated leptons                                  & 0.41 & 0.50 \\
$80 \; \text{GeV} < M_{\ell\ell} < 100$ GeV, $p_{T,\ell\ell} > 160$ GeV, $\Delta R_{\ell\ell} > 0.2$ & 0.83 & 0.89 \\
At least 1 fat jet with 2 $B$-meson tracks with $p_T > 110$ GeV & 0.96 & 0.98 \\
2 Mass drop subjets and $\ge 2$ filtered subjets & 0.88 & 0.92 \\
2 $b$-tagged subjets & 0.38 & 0.41 \\
$115 \; \text{GeV} < m_h < 135$ GeV & 0.15 & 0.51 \\
$\Delta R(b_i, \ell_j) > 0.4$, $\slashed{E}_T < 30$ GeV, $|y_h| < 2.5$, $p_{T,h/Z} > 200$ GeV & 0.47 & 0.69 \\   
\hline 
 \end{tabular}
\caption{Cut-flow table showing the effect of each cut on $Zb\bar{b}$ and SM $Zh$.}
\label{tab:cut-flow}
\end{center}
\end{table*}

\subsection{Analysis}
In order to probe the reach of the high luminosity runs of the LHC in constraining the EFT directions in Tab.~\ref{dirn}, we 
optimize a hadron-level analysis to obtain maximum sensitivity to the BSM signal, which is well-pronounced in the 
high energy bins. To achieve this, we consider the $Z(\ell^+\ell^-)h$ production from a pair of quarks as well as from a 
pair of gluons. As far as the decay of the Higgs boson is concerned,  we find that at an integrated luminosity of  300 fb$^{-1}$,  the diphoton mode yields less than 5 events at high energies ($p_{T,Z}>150$ GeV) and is thus not sensitive to the effects we want to probe. We thus scrutinize the $h(b\bar{b})Z(\ell^+\ell^-)$ final state  where the dominant backgrounds are composed of
$Zb\bar{b}$ and the irreducible SM production of $Zh$. For the $Zb\bar{b}$ process, 
we consider the tree-level production as well as the $gg \to ZZ$ production at one-loop.
Reducible contributions arise from $Z+$ jets production ($c$-quarks included but not explicitly
tagged), where the light jets can be misidentified as $b$-jets, and the fully leptonic decay for 
$t\bar{t}$. Instead of performing a standard resolved analysis, where one would demand two separate
$b$-tagged jets, we demand a fat jet with a cone radius of $R=1.2$. We employ the so-called BDRS 
approach~\cite{Butterworth:2008iy} with minor modifications to maximize sensitivity. In a nutshell, 
this technique helps in discriminating boosted electroweak-scale resonances from large QCD backgrounds.  

We will see that using this approach will allow us to reduce the ratio of  $Zbb$ to SM $Zh$ events from about 40 to an ${\cal O}(1)$ number with about 40 SM events still surviving at 300 fb$^{-1}$. This shows that the kind of analysis performed here would not be possible at integrated luminosities smaller than  300 fb$^{-1}$.

The BDRS approach recombines jets using the Cambridge-Aachen (CA) algorithm~\cite{Dokshitzer:1997in,Wobisch:1998wt} with a significantly large
cone radius to contain all the decay products of the resonance. One then works backwards
through the jet clustering and stops when a significant {\emph{mass drop}}, 
$m_{j_1} < \mu m_j$ with $\mu=0.66$, ($m_j$ being the mass of the fatjet) occurs for a not too 
asymmetric splitting
, 
$$\frac{\text{min}(p_{T,j_1}^2,p_{T,j_2}^2)}{m_j^2}\Delta R_{j_1,j_2}^2>y_{\text{cut}},$$ 
with
$y_{\text{cut}} = 0.09$. If this condition is not met, the softer subjet, $j_2$ is removed and the subjets of $j_1$ are tested
for the aforementioned criteria to be satisfied in an iterative process. The algorithm stops as soon as one obtains two subjets,
$j_1$ and $j_2$ abiding by the mass drop condition. 

To improve the resonance reconstruction, the technique considers a further step called \textit{filtering}. In this
step, the constituents of $j_1$ and $j_2$ are again recombined using the CA algorithm with a cone radius
$R_{\text{filt}} = \text{min}(0.3, R_{b\bar{b}}/2)$. Only the hardest three filtered subjets are retained
to reconstruct the resonance. In the original work of Ref.~\cite{Butterworth:2008iy}, this resonance is the SM-like Higgs and thus the two
hardest filtered subjets are $b$-tagged. In our work, we find that the filtered cone radius 
$R_{\text{filt}} = \text{max}(0.2, R_{b\bar{b}}/2)$ works better in removing the backgrounds.\footnote{The criteria $R_{\text{filt}} = \text{max}(0.2, R_{b\bar{b}}/2)$ followed by $R_{\text{filt}} = 
\text{min}(0.3, R_{b\bar{b}}/2)$ hardly changes the results.} The filtering step greatly reduces the active area
of the initial fatjet. 

We use the \texttt{FeynRules}~\cite{Alloul:2013bka} and \texttt{UFO}~\cite{Degrande:2011ua} toolkits to implement the signal contributions (we will comment on the effect of including squared dimension 6 interactions as compared to interference-only terms below).
Both signal and background processes are generated including the full decay chain with \texttt{MG5$\_$aMC@NLO}~\cite{Alwall:2014hca}, at leading order.
For the gluon initiated part of the SM and BSM $Zh$ production, we employ the \texttt{FeynArts/FormCalc/LoopTools}~\cite{Hahn:2000kx,Hahn:1998yk} 
framework and the decays are performed using \texttt{MadSpin}~\cite{Frixione:2007zp,Artoisenet:2012st}. We shower and hadronize the samples using \texttt{Pythia 8}~\cite{Sjostrand:2001yu,Sjostrand:2014zea} and 
perform a simplified detector analysis. 

Because our ultimate goal is to look for new physics effects in 
high energy bins, we generate the $Zh$, $Zb\bar{b}$ and $t\bar{t}$ samples with the following cuts: 
$p_{T,(j,b)} > 15$ GeV, $p_{T,\ell} > 5$ GeV, $|y_j| < 4$, $|y_{b/\ell}| < 3$, $\Delta R_{bb/bj/bl} > 0.2$, 
$\Delta R_{\ell\ell} > 0.15$, $70 \; \text{GeV} < m_{\ell \ell} < 110$ GeV, $75 \; \text{GeV} < m_{bb} < 
155$ GeV and $p_{T,\ell\ell} > 150$ GeV. The former two processes are generated upon merging with an 
additional matrix element (ME) parton upon using the MLM scheme~\cite{Mangano:2006rw}. 
For the $Z+$jets process, we generate the samples without the invariant mass cuts on the jets; we 
further merge the sample up to three ME partons. 

To account for higher order QCD corrections for the $q\bar{q}$-initiated $Zh$ 
process, we apply a bin-by-bin (in $M_{Zh}$, the invariant mass of the filtered double $b$-tagged fat jet and the reconstructed $Z$-boson) $K$-factor reweighting to the NLO-accurate distribution both for the SM background and the EFT signal using Ref.~\cite{Greljo:2017spw}. For 
the $gg$ initiated $Zh$ process, we consider a conservative NLO $K$-factor of 2~\cite{Altenkamp:2012sx}. 
For the tree-level $Zb\bar{b}$ and $Z+$jets processes, flat $K$-factors of 1.4 (computed within \texttt{MG5$\_$aMC@NLO}) and 0.91~\cite{Campbell:2002tg} are applied.
For the $gg \to ZZ$ production, a flat $K$-factor of $\sim1.8$~\cite{Alioli:2016xab} has been
used.

We first test the power of a cut-based analysis. In doing so, we construct
the fatjets with a cone radius of $R=1.2$, having $p_T > 80$ GeV and rapidity, $|y| < 2.5$ using 
\texttt{FastJet}~\cite{Cacciari:2011ma}. We isolate the leptons ($e,\mu$) by demanding that the total hadronic activity around a cone radius of $R=0.3$ must be less than
10\% of its $p_T$ and the leptons are required to have $p_T > 20$ GeV and $|y| < 2.5$. All non-isolated objects
are considered while constructing the fatjets. In selecting our events, we consider only those with
exactly two isolated leptons having opposite charge and same flavour (OSSF). Moreover, we demand the 
invariant mass of the pair of leptons to lie in the range $[80~\text{GeV},100~\text{GeV}]$ in order to reconstruct the 
$Z$-peak. The reconstructed $Z$ is required to be boosted with $p_T > 160$ GeV and the separation between
the two isolated leptons is required to be $\Delta R > 0.2$. In reconstructing the Higgs boson, we demand
that each event has at least one fatjet containing no less than two $B$-meson tracks with $p_T > 15$ GeV.
The minimum transverse momentum of the fatjet is required to be $p_T > 110$ GeV. After satisfying the 
mass drop and filtering criteria, we require exactly two subjets after the former step
and at least two subjets after filtering. We proceed with $b$-tagging the two hardest 
subjets. We choose a $b$-tagging efficiency of 70\% and a misidentification rate for light 
jets of 2\%. After the filtering and $b$-tagging steps, we require events with exactly two $b$-tagged 
subjets, which are well-separated from the isolated leptons: 
$\Delta R(b_i,\ell_j) > 0.4$ for both leptons $\ell_{1,2}$ and $b$-tagged subjets $b_i$. 
We reconstruct the Higgs by requiring its invariant mass to lie in the range [115 GeV, 135 GeV]. 

In order to further reduce the backgrounds, we demand both the reconstructed $Z$ and the Higgs
bosons to have $p_T > 200$ GeV. The $t\bar{t}$ background can be removed almost entirely by requiring 
$\slashed{E}_T < 30$ GeV. The cut-flow affecting the most dominant background $Zb\bar{b}$ and the SM 
$Zh$ channel, is summarized in Table~\ref{tab:cut-flow}. 

Before focussing on the very high-energy effects by imposing
cuts on $M_{Zh}$, we find that the ratio of cross-section between SM $Zh$ and $Zb\bar{b}$ is $\sim 0.26$.
A multivariate implementation at this level strengthens this ratio further. In order
to be quantitative, we impose looser cuts on the aforementioned variables $70 \; \text{GeV} < 
m_{\ell\ell} < 110$ GeV, $p_{T,\ell\ell} > 160$ GeV, $\Delta R_{\ell\ell} > 0.2$, 
$p_{T,\text{fatjet}} > 60$ GeV, $95 \; \text{GeV} < m_h < 155$ GeV, $\Delta R_{b_i,\ell_j} > 0.4$ and 
$\slashed{E}_T < 30$ GeV. Because $Z+$jets and $t\bar{t}$ are much less significant than $Zb\bar{b}$, we
train the boosted decision trees only with the SM $q\bar{q}$-initiated $Zh$ and $Zb\bar{b}$ samples 
using the following variables: $p_T$ of the two isolated leptons, $\Delta R$ between pairs of $b$-subjets 
and isolated leptons, between the two isolated leptons and between the hardest two $b$-subjets in the Higgs 
fatjet, the reconstructed $Z$-boson mass and its $p_T$, $\Delta \Phi$ separation
between the fatjet and the reconstructed $Z$-boson, $\slashed{E}_T$, mass of the reconstructed Higgs jet
and its $p_T$, $p_T$ of the two $b$-tagged filtered subjets, the ratio of their $p_T$ and the rapidity of
the Higgs jet. We ensure that we do not have variables which are $\sim100\%$ correlated but we retain all 
other variables. Because our final distribution of interest is the invariant mass of the $Zh$-system, we
do not consider it as an input variable. We use the \texttt{TMVA}~\cite{2007physics3039H} framework to
train our samples and always ensure that the Kolmogorov-Smirnov statistic is at least of the order $\sim0.1$
in order to avoid overtraining of the samples~\cite{KS}. We find that the aforementioned ratio increases
to $\sim 0.50$ upon using the boosted decision tree algorithm showing that a further optimisation of the cut-based
analysis was necessary. Finally, we test all our samples with the training obtained from the SM $q\bar{q}$
initiated $Zh$ and the $Zb\bar{b}$ samples.

\begin{figure}[!t]
\includegraphics[scale=0.5]{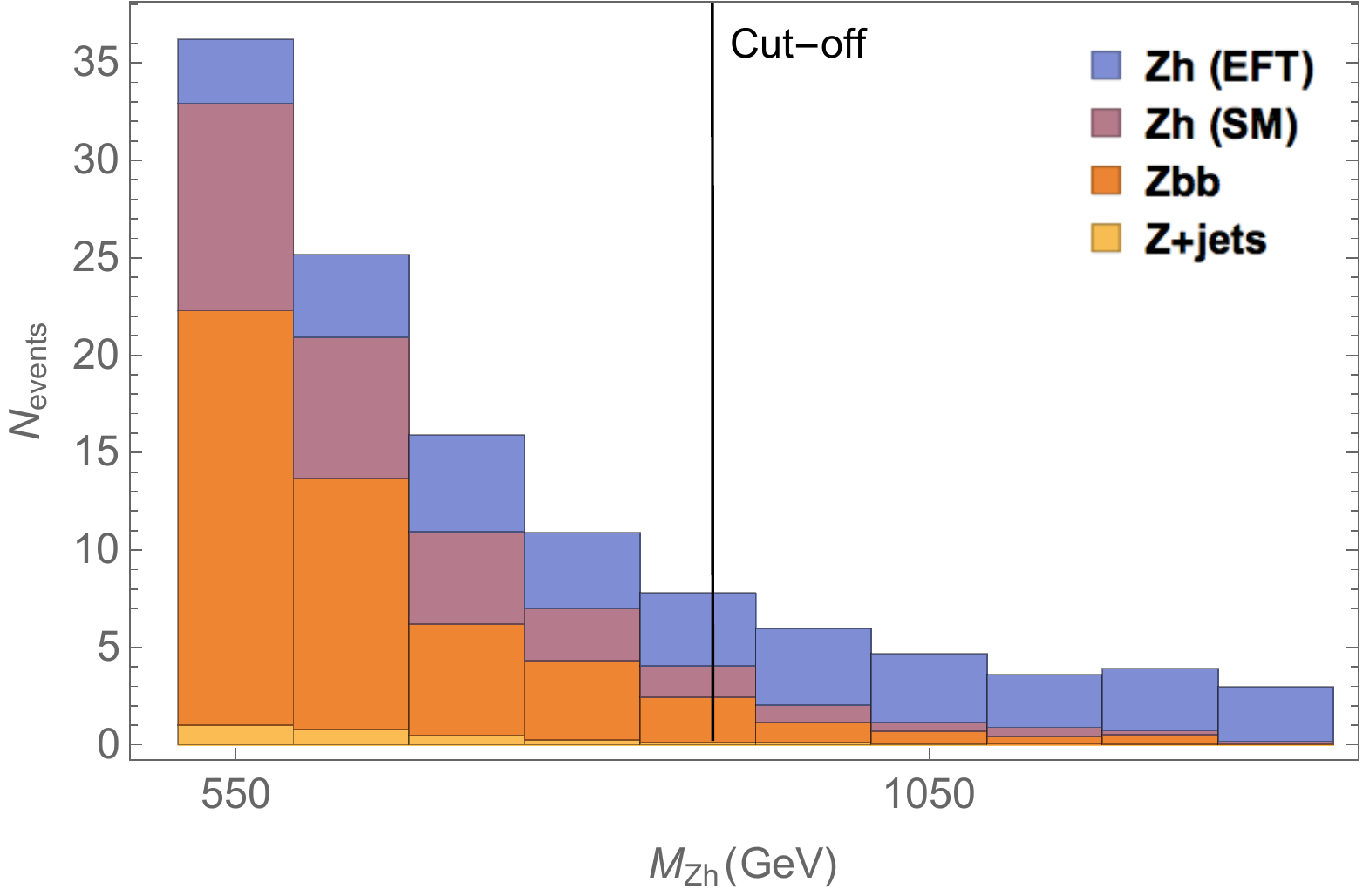}
\caption{The differential distribution of events at an integrated luminosity of 300 fb$^{-1}$ with respect to $M_{Zh}$ for the EFT signal as well as the different backgrounds. For the EFT signal we have taken the point $\{{g}^h_{Zu_L} ,{g}^h_{Zd_L} ,{g}^h_{Zu_R} ,{g}^h_{Zd_R} \}=\{-0.005, 0.0001 , -0.010, 0.005 \}$ which is allowed by the LEP bounds.}
\label{bars}
\end{figure}

To distinguish between the EFT signal and the irreducible SM $Zh(b\bar b)$ background we utilise the 
growth of the EFT cross-section at high energies. The effects are readily seen in the $M_{Zh}$ distribution, 
our observable of interest. In Fig.~\ref{bars} we show the differential distribution with respect to 
this variable for the EFT signal as well as the different backgrounds for an integrated luminosity of 
300 fb$^{-1}$. For the EFT signal we take a point that can be excluded in our analysis but is well within the LEP allowed region. We see that the 
EFT cross-section keeps growing with energy, but much of this growth is unphysical at energies above 
the cut-off, \textit{i.e.}, $M_{Zh} >\Lambda$, where $\Lambda$ is the cut-off evaluated as described 
below \eq{cutoff} and shown by a vertical line in Fig.~\ref{bars}. For our analysis we dropped all events above this cut-off. For $M_{Zh} <\Lambda$, the EFT 
deviations are never larger than an ${\cal O}(1)$ factor with respect to the SM background as 
expected on general grounds. Note, however, that even for $M_{Zh} <\Lambda$, even though 
the underlying anomalous couplings, $g^h_{Zf}$, are per-mille to percent level, the 
fractional deviations are much larger because of the energy growth of the BSM rate. To 
make full use of the shape deviation of the EFT signal with respect to the background, 
we perform a binned log likelihood analysis assuming a 5$\%$ systematic error. The 
likelihood function is taken to be the product of Poisson distribution functions for 
each bin with the mean given by the number of events expected for a given BSM point. 
To account for the 5$\%$ systematic error we smear the mean with a Gaussian 
distribution. To obtain the projection for the 95$\%$ CL exclusion curve we assume that 
the observed number of events agrees with the SM.
\subsection{Discussion}

  

\begin{table}[t]
\begin{center}

\begin{tabular}{c|c|c}
&Our Projection &LEP Bound\\\hline
$\delta g^Z_{u_L}$         & $\pm0.002~(\pm0.0007)$ & $-0.0026\pm 0.0016$\\
$\delta g^Z_{d_L}$         & $\pm0.003~(\pm0.001)$  & $0.0023\pm 0.001$\\
$\delta g^Z_{u_R}$         & $\pm0.005~(\pm0.001)$  & $-0.0036\pm 0.0035$\\
$\delta g^Z_{d_R}$         & $\pm0.016~(\pm0.005)$  & $0.016\pm 0.0052$\\
$\delta g^Z_1$             & $\pm0.005~(\pm0.001)$  & $0.009^{+0.043}_{-0.042}$\\
$\delta \kappa_\gamma$     & $\pm0.032~(\pm0.009)$  & $0.016^{+0.085}_{-0.096}$\\
$\hat{S}$                  & $\pm0.032~(\pm0.009)$ & $0.0004 \pm 0.0007$\\
$W$                        & $\pm0.003~(\pm0.001)$  & $0.0000 \pm 0.0006$\\
$Y$                        & $\pm0.032~(\pm0.009)$  & $0.0003\pm 0.0006$
 \end{tabular}
  \caption{Comparison of the bounds obtained in this work with existing LEP bounds. The numbers outside (inside) brackets, 
  in the second column, denote our bounds with $\mathcal{L} = 300 \; (3000)$ fb$^{-1}$. To obtain our projection we turn on the LEP observables in \eq{diru} one by one and use \eq{obo}. The LEP bounds on the $Z$ coupling to quarks has been obtained from Ref.~\cite{Falkowski:2014tna}, the bound on the TGCs from Ref.~\cite{LEP2}, the bound on $\hat{S}$ from Ref.~\cite{Baak:2012kk} and  finally the bounds on $W,Y$ have been obtained from Ref.~\cite{Barbieri:2004qk}. Except for the case of the bounds on $\delta g^Z_f$, all of the bounds in the last column were derived by turning on only the given parameter and putting all other parameters  to zero. \label{lepb}}
\end{center}
\end{table}
\begin{figure}[!b]
\includegraphics[scale=0.6]{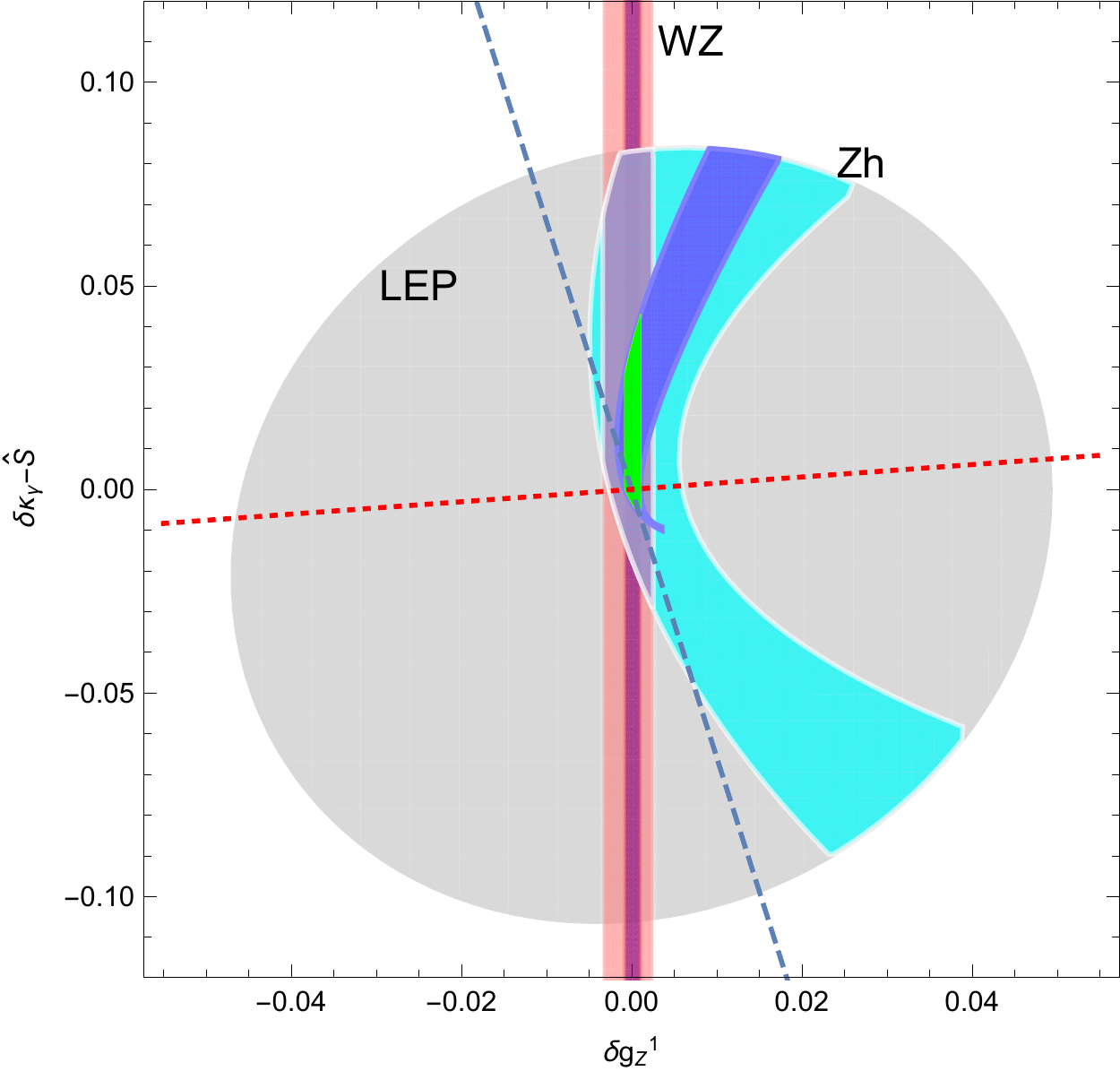}
\caption{We show in light blue (dark blue)  the projection for the allowed region with 300 fb$^{-1}$ (3 ab$^{-1}$) data from the 
$pp \to Zh$ process for universal models in the $\delta \kappa_\gamma-\hat{S}$ vs $\delta g^Z_1$ plane. The allowed region after LEP bounds (taking the TGC $\lambda_\gamma=0$, a conservative choice) are imposed is shown in grey.  The pink (dark pink)  region corresponds to the projection from the $WZ$ process  with 300 fb$^{-1}$ (3 ab$^{-1}$) data derived in Ref.~\cite{Franceschini:2017xkh} and the purple (green) region shows the region that survives after our projection from the $Zh$ process is combined with the above $WZ$ projections with 300 fb$^{-1}$ (3 ab$^{-1}$) data.}\label{bounds}
\end{figure} 
Considering only the SM-BSM interference term, we find the per-mille level bounds, 
\begin{eqnarray}
g^h_{Z\textbf{p}} &\in & \left[-0.004,0.004\right]~~~~(300{\rm~fb}^{-1}) \nonumber\\
g^h_{Z\textbf{p}} &\in & \left[-0.001,0.001\right]~~~~(3000{\rm~fb}^{-1}).
\label{obo}
\end{eqnarray}
Using \eq{cutoff} the above bounds  can be translated to a lower bound on the scale of new physics given by 2.4 TeV (4.4 TeV) at 300 fb$^{-1}$ (3000 fb$^{-1}$).  One can now compare the above projections with existing LEP bounds  by  turning on the LEP observables contributing to $g^h_{Z\textbf{p}}$ in \eq{diru}  one by one. This is equivalent to assuming that there are no large cancellations in \eq{diru} so that each individual term   is bounded by \eq{obo}.  The results are shown in Tab.~\ref{lepb}. We see that our projections are much stronger than the LEP bounds for the TGCs $\delta g^Z_1$ and $\delta \kappa_\gamma$ and comparable in the case of the $Z$-pole observables $\delta g^Z_f$, that parametrize the deviations of the $Z$ coupling to quarks. 

To obtain \eq{obo}, we have used the cut-off as defined in \eq{cutoff} with $g_h=1$. One might expect a stronger bound by taking a larger $g_h$. We find, however, that while taking stronger couplings can increase the cut-off many times, this does not lead to an appreciably  higher sensitivity because the high energy bins have very few or no SM/BSM events being suppressed by the small PDFs at these energies.

For the universal case, the EFT directions presented in Table~\ref{dirn} can be visualized in the 
$\delta \kappa_\gamma-\hat{S}$ vs. $\delta g^Z_1$ plane as shown in Fig.~\ref{bounds} for the interesting class of models where $W=Y=0$~\cite{Franceschini:2017xkh}. The flat 
direction related to the $pp \to Zh$ interference term, \textit{i.e.}, $g^h_{Z\textbf{p}}=0$, 
\eq{compdir}, is shown by the dashed blue line, where the direction $g^h_{Z\textbf{p}}$ is now given 
by the second line of  \eq{diru}. The grey shaded area shows the allowed region after the LEP II bounds~\cite{LEP2} from the $e^+e^- \to W^+W^-$ 
process are imposed. The results of this work are shown in blue (light (dark) blue for results at
300 (3000) fb$^{-1}$). To understand the shape of the blue bands, note that along the dashed line, 
the SM-BSM interference term vanishes. If the interference was the only dominant effect, the 
projected allowed region would be a band along this direction. The BSM squared term thus plays a 
role in determining the shape of the blue region. To the left of the dashed blue line, the squared 
and the interference terms have the same sign while there is a partial cancellation between these 
two terms on the right hand side of the dashed line. This results in the curvature of the blue band 
with stronger bounds to the left of the dashed line and weaker bounds to its right. 

We see that, as we move further from the origin, the effect of the squared term becomes more pronounced. This is expected, as along the dashed line, the interference term is accidentally zero, even for energies below the cut-off, and thus, the parametrically sub-dominant squared term is larger. To 
achieve a partial cancellation between these two terms one needs to 
deviate more and more from the dashed line. While EFT validity has been carefully imposed to derive 
our bounds, the fact that the interference term vanishes along the flat direction and the squared 
term becomes important, does imply that for weakly coupled UV completions our bounds are susceptible 
to ${\cal O}(1)$ dimension 8 deformations in this direction. In the orthogonal direction shown by the 
dotted line, on the other hand, our projections are more robust and not sensitive to such effects. Such an ambiguity also exists in the results  in Tab.~\ref{lepb}, for the pseudo-observables such as $\delta g^Z_{d_R}$ and $\delta \kappa_\gamma$, that are somewhat aligned to the above flat direction. This ambiguity can be resolved by performing a global fit upon combining analyses of all the $Vh, VV$ channels, that will avoid such flat directions.
  
As we have emphasized already, $VV$ production constrains the same set of operators as the $Vh$ 
production. In Fig.~\ref{bounds}, we also show the projected bound from the $WZ$ process at 300 
fb$^{-1}$ obtained in Ref.~\cite{Franceschini:2017xkh}. When both these bounds are combined, only the 
purple region remains. At 3000 fb$^{-1}$, this region shrinks further to the green region shown in 
Fig.~\ref{bounds}. Thus, we see a drastic reduction in the allowed LEP region is  possible 
by investigating $pp \to Zh$ at high energies.

\subsection{Conclusions}
As hints for new physics beyond the SM remain elusive with the LHC entering a new energy territory, model-independent approaches based on the assumption of no additional light propagating degrees of freedom are gaining ground. The power of effective field theory is that theoretical correlations between independent measurements can be exploited to formulate tight constraints on the presence of new physics, solely based on the SM symmetries and matter content.

The high precision measurements performed during the LEP era are therefore the driving forces behind combined constraints early in the LHC program. To enter new territory, the LHC has to push beyond the LEP sensitivity for interactions that relate the phenomenology at both collider experiments. The Higgs boson, as arguably the most significant TeV scale degree of freedom, can be placed at the core of such a program, that will naturally involve LHC measurements at high luminosity. 

In this paper, we focussed on the impact of associated Higgs production that provides complementary information to the diboson production modes observed at LEP2, which determine the precision of the associated coupling constraints. Using a dedicated investigation of expected signal and backgrounds, we find that the LHC will ultimately be able to improve the sensitivity expected from LEP measurements. Our results are summarised in \eq{obo}, Tab.~\ref{lepb} and Fig.~\ref{bounds}. Higgs-strahlung is also complementary to diboson production at LHC  investigated in Ref.~\cite{Franceschini:2017xkh}.  Combining Higgs-strahlung measurements with diboson results in the high energy limit will allow us to drastically improve the sensitivity to the underlying new physics parameters in an unparalleled way.

Both high energies and luminosities are crucial for a study like ours. Potentially even higher new physics scales can thus be probed at the High Energy LHC or other future colliders.
        
\subsection{Acknowledgements}
We thank Matthew~McCullough for helpful discussions and collaboration on this project in its early stage.
We also thank Biplob Bhattacherjee, Mikael Chala, Shilpi Jain, Giuliano Panico, Michael Peskin and Francesco Riva for 
helpful discussions at various stages of this work.

C.E. is supported by the IPPP Associateship scheme and by the UK Science and Technology Facilities Council (STFC) under grant ST/P000746/1.
S.B. is supported by a Durham Junior Research Fellowship COFUNDed by Durham University and the 
European Union, under grant agreement number 609412.

\bibliography{references}    

\end{document}